\documentclass[conference]{IEEEtran}
\IEEEoverridecommandlockouts

\usepackage{cite}
\usepackage{comment}
\usepackage{amsmath,amssymb,amsfonts}
\usepackage{algorithmic}
\usepackage{graphicx}
\usepackage{textcomp}
\usepackage{balance}
\usepackage{tabularx}
\usepackage{makecell}
\usepackage{multirow}

\usepackage{subcaption} 
\usepackage[table,xcdraw]{xcolor}
\DeclareMathOperator{\argmin}{argmin}
\def\BibTeX{{\rm B\kern-.05em{\sc i\kern-.025em b}\kern-.08em
    T\kern-.1667em\lower.7ex\hbox{E}\kern-.125emX}}
\begin{document}

\title{Performance Analysis of Spatial and Temporal Learning Networks in the Presence of DVL Noise}

\author{Rajini Makam$^{1*\dag}$~\IEEEmembership{Member,~IEEE},  Nadav Cohen$^{2\dag}$,~\IEEEmembership{Graduate Student Member,~IEEE},  Sumukh Shadakshari$^{3}$,\\ Srinivasa Puranika Bhatta$^{3}$,  Itzik Klein$^{2}$,~\IEEEmembership{Senior Member,~IEEE}, and Suresh Sundaram$^{1}$~\IEEEmembership{Senior Member,~IEEE}
\thanks{$^{*}$Corresponding Author, $^{\dag}$Equal Contribution}%
\thanks{$^{1}$Rajini Makam \& Suresh Sundaram are with the Department of Aerospace Engineering, Indian Institute of Science, Bangalore, India.
        {\tt\small \{rajinimakam@iisc.ac.in \& vssuresh@iisc.ac.in\}}}%
\thanks{$^{2}$ Nadav Cohen \& Itzik Klein are with The Hatter Department of Marine Technologies, Charney School of Marine Sciences, University of Haifa, Haifa, Israel.
        {\tt\small \{ncohe140@campus.haifa.ac.il \& kitzik@univ.haifa.ac.il\}}}%
\thanks{$^{3}$ Sumukh Shadakshari \& Srinivasa Puranika Bhatta are with the Dept. of Elect. and Comm. Engn., PES University, Bangalore, India. {\tt\small \{sumukhshad19@gmail.com, sp.bhatta@puranika.com\}}}%
}

\maketitle

\begin{abstract}
Navigation is a critical aspect of autonomous underwater vehicles (AUVs) operating in complex underwater environments. Since global navigation satellite system (GNSS) signals are unavailable underwater, navigation relies on inertial sensing, which tends to accumulate errors over time. To mitigate this, the Doppler velocity log (DVL) plays a crucial role in determining navigation accuracy. In this paper, we compare two neural network models: an adapted version of BeamsNet, based on a one-dimensional convolutional neural network, and a Spectrally Normalized Memory Neural Network (SNMNN). The former focuses on extracting spatial features, while the latter leverages memory and temporal features to provide more accurate velocity estimates while handling biased and noisy DVL data. The proposed approaches were trained and tested on real AUV data collected in the Mediterranean Sea. Both models are evaluated in terms of accuracy and estimation certainty and are benchmarked against the least squares (LS) method, the current model-based approach. The results show that the neural network models achieve over a 50\%  improvement in RMSE for the estimation of the AUV velocity, with a smaller standard deviation.
\end{abstract}

\section{Introduction}

Autonomous Underwater Vehicles are essential for underwater applications such as exploration, environmental monitoring, and naval operations. Accurate velocity estimation is crucial in enabling precise navigation and control of AUVs. Although inertial navigation systems (INS) provide navigation data, noise causes orientation, velocity, and position drift over time. Sensors such as the GNSS, DVL and pressure sensors (PS) correct this by providing position, velocity, and height data \cite{Mu2021, Wang2020}. Inertial measurement units (IMUs) offer critical data on specific force and angular velocities. When combined with DVL velocity measurements, this allows for an accurate navigation solution, specifically underwater, which is a GNSS-denied environment with environmental disturbances \cite{groves2015principles,Taudien2018}.



To improve navigation accuracy, DVL was integrated with IMU data using traditional fusion algorithms such as the extended Kalman filter (EKF), unscented Kalman filter (UKF), and Particle Filter (PF) \cite{karimi2013comparison,Seamless24, Chi2021}. As a result, machine learning models have gained attention for their ability to capture intricate patterns in sensor data \cite{zhang2023}.  In recent years, Deep Neural Networks (DNNs) have gained increasing importance, particularly when large datasets from extended missions are available. This has led to improvements in inertial navigation in the land, aerial, and maritime domains \cite{cohen2023inertial}. In \cite{zhang2020}, a fault-tolerant deep learning model called "NavNet" was introduced, which integrates the Attitude and Heading Reference System (AHRS) with DVL for more precise navigation estimates. The approach was further advanced with the use of a recurrent neural network (RNN) in \cite{mu2019}. In \cite{topini2020}, the authors proposed a combined dead reckoning and long short-term memory (LSTM) method. A radial basis function is given in \cite{liu2022} to correct for current-induced errors in INS and DVL. In \cite{topini2023}, a combination of CNN-LSTM called C-LSTM is proposed to predict AUV pose.

In this paper, we compare these two models, highlighting their complementary strengths in addressing different aspects of the data. BeamsNet, which is based on 1D-CNN, excels at learning spatial features, making them highly effective in identifying patterns that could indicate the AUV’s velocity at a specific moment \cite{cohen2022}. On the other hand, memory-augmented neural networks focus on capturing temporal dynamics, enabling them to account for the sequence of velocity changes and how past states influence current behavior \cite{Makam2024}. In the presence of bias and noise, this comparison is essential to determining whether CNNs, which extract spatial information, or memory networks, which maintain predictive stability, are superior in terms of accuracy, complexity, and storage. The performance of this algorithm is evaluated against the least squares (LS) method.

The rest of the paper is organized as follows: Section \ref{PF} formulates the task at hand, Section \ref{meth} presents data-driven methodologies, Section \ref{RAD} describes data acquisition and discusses the results, and Section \ref{con} concludes the findings.

\section{Problem Formulation} \label{PF}

The DVL is used to measure the AUV's velocity, thereby determining the accuracy of the navigation solution. To determine the velocity, the DVL transmits four acoustic beams in an `$\times$'-shaped configuration, known as the Janus Doppler configuration. In this configuration, the transducers are positioned at specific yaw and pitch angles relative to the DVL's sensor body frame, as follows:
\begin{equation}\label{eqn:1}
    \centering
        \boldsymbol{b}_{\dot{\imath}}=
        \begin{bmatrix} 
        \cos{\Psi_{\dot{\imath}}}\sin{\Theta}\quad
        \sin{\Psi_{\dot{\imath}}}\sin{\Theta}\quad
        \cos{\Theta}
    \end{bmatrix}_{1\times3}, \dot{\imath} = 1, 2, 3, 4.
\end{equation} 
where, $\boldsymbol{b}_{\dot{\imath}}$ represents the beam number, $\Psi$ and $\Theta$ correspond to the yaw and pitch angles relative to the body frame. The pitch angle is fixed and pre-determined by the manufacturer, ensuring the same value for all beams, while the yaw angle is defined as:
\begin{equation}\label{eqn:2} 
\centering \Psi_{\dot{\imath}} = (\dot{\imath} - 1) \cdot 90^{\circ} + 45^{\circ}. 
\end{equation}
By defining a transformation matrix $\mathbf{H}$, the relationship between the DVL's velocity in the body frame, $\boldsymbol{v}_{b}^{b}$, and the beam velocity measurements, $\boldsymbol{v}^{beam}$, can be expressed as:
\begin{equation}\label{eqn:3} 
\centering 
    \boldsymbol{v}^{beam} = \mathbf{H} \boldsymbol{v}_{b}^{b}, \quad \mathbf{H} =         \begin{bmatrix} 
        \boldsymbol{b}_{1}\\\boldsymbol{b}_{2}\\\boldsymbol{b}_{3}\\\boldsymbol{b}_{4}\\
    \end{bmatrix}_{4\times3}.
\end{equation}
The beam velocity vector, $\boldsymbol{v}^{beam}$, represents the raw output of the DVL and is typically affected by in-run bias and white processing noise, as described by the following error model:
\begin{equation}\label{eqn:4}
    \centering
        \tilde{\boldsymbol{v}}^{beam}= \mathbf{H}\boldsymbol{v}_{b}^{b}+\boldsymbol{b}_{DVL}+\boldsymbol{n},
\end{equation}
where $\boldsymbol{b}_{DVL}$ is the in-run bias represented as a $4\times1$ vector, $\boldsymbol{n}$ denotes zero-mean white Gaussian noise, and $\tilde{\boldsymbol{v}}^{beam}$ represents the corrupted beam velocity measurements.\\
To extract the AUV velocity from the DVL's raw data, the literature recommends using a least-squares (LS) formulation:
\begin{equation}\label{eqn:5}
    \centering
        \hat{\boldsymbol{v}}_{b}^{b}=
        \underset{\boldsymbol{v}_{b}^{b}}{\argmin}{\mid\mid\tilde{\boldsymbol{v}}^{beam}-\mathbf{H}\boldsymbol{v}_{b}^{b} \mid\mid}^{2}.
\end{equation} 
The solution for $\hat{\boldsymbol{v}}_{b}^{b}$ is obtained by multiplying the pseudoinverse of the transformation matrix $\mathbf{H}$ by the raw beam measurements:
\begin{equation}\label{eqn:6}
    \centering
        \hat{\boldsymbol{v}}_{b}^{b}= H^{\dag} \tilde{\boldsymbol{v}}^{beam},
\end{equation} 
where, $H^{\dag} = (\mathbf{H}^{T}\mathbf{H})^{-1}\mathbf{H}^{T} $. The underlying assumption regarding the LS method in this context is that it represents the best linear estimator. Furthermore, when the estimator is unbiased and the measurements are corrupted by independent and identically distributed Gaussian noise with known variance, the LS estimator is efficient and achieves the Cramér-Rao Lower Bound, which means that its variance is the lowest possible for an unbiased estimator \cite{bar2004estimation}. In real-world applications, calibration may not completely eliminate in-run bias, and the noise may not be strictly white. As a result, these assumptions are often not fully satisfied, suggesting that more effective estimators could be available. 

\section{Methodology} \label{meth}

In this section, we present two learning strategies one based on the convolutional neural network (CNN) and the other on memory-based network. 

\subsection{BeamsNet}
\begin{figure*}[h!]
	\centering
		\includegraphics[scale=0.25]{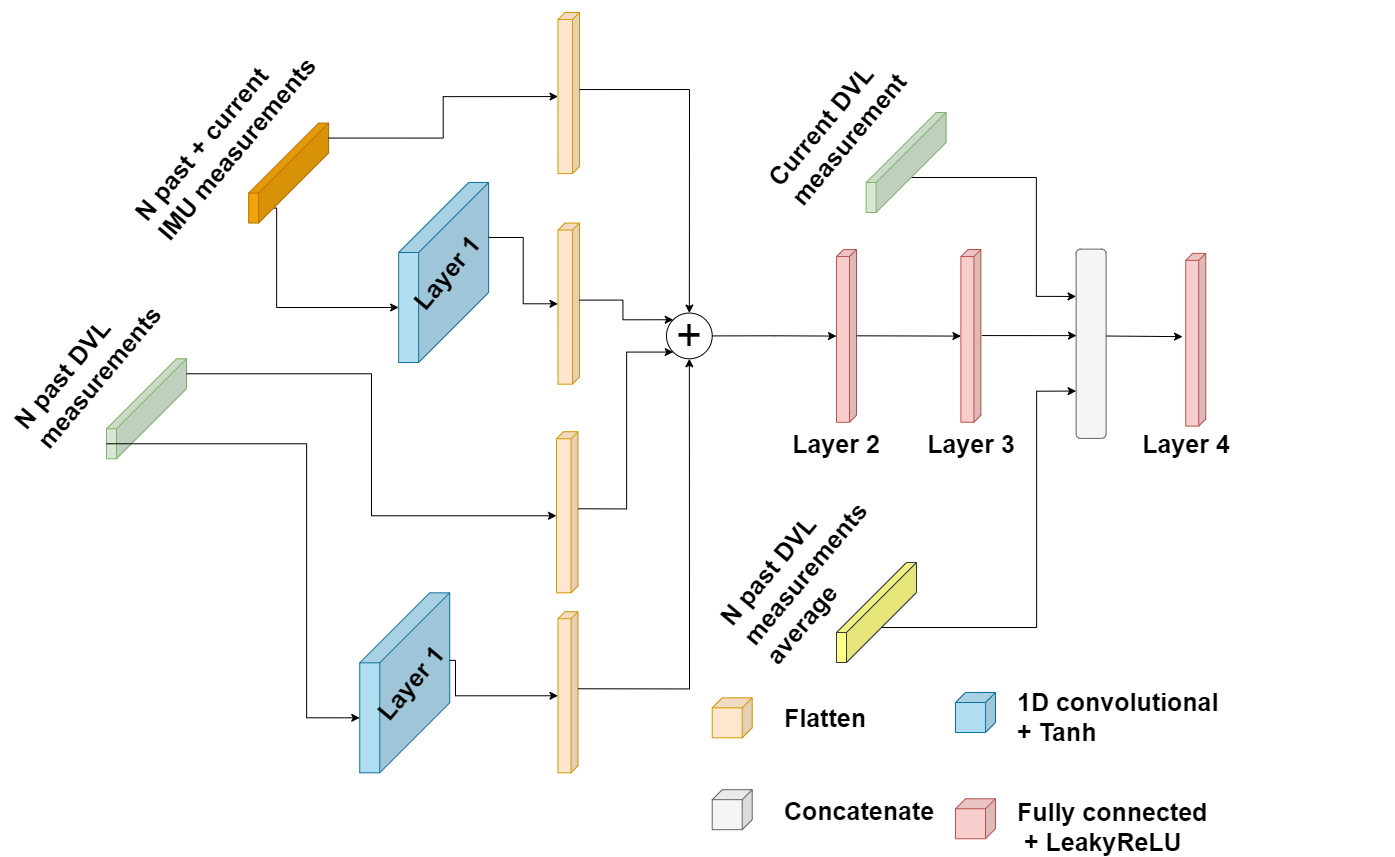}
	  \caption{The modified BeamsNet architectures: BeamsNetV1 and BeamsNetV2}\label{fig1}
\end{figure*}

The first methodology examined is a modified BeamsNet framework, which has two variations. In the first version, past DVL beam measurements and IMU data were used as input to a 1D CNN-based network to predict the AUV velocity. The second version assumes that no IMU data are available, using only past DVL beam measurements as input \cite{cohen2022}. The fisrt version of modified BeamsNet (BeamsNetV1) architecture can be seen in Fig. \ref{fig1}. The first layer of the network consists of two 1D convolutional layers that process both past and current IMU and DVL data. After a residual connection, the features extracted from the IMU and DVL are merged and passed through a series of fully connected layers. The result is then concatenated with the average of the previous DVL measurements along with the current DVL measurement before passing through a final fully connected layer to predict the AUV velocity. In the second version of the modified BeamsNet (BeamsNetV2), which only uses DVL data, all blocks related to the IMU are removed, and the hyperparameters are adjusted accordingly. The hyperparameters are summarized in Table \ref{tab:1}. For both versions, the loss function is the mean squared error (MSE), and the ADAM optimizer is applied with a learning rate of 0.01 and a batch size of 256. A LeakyReLU activation function with a negative slope of 0.1 follows each fully connected layer. BeamsNetV1 was trained for 250 epochs with a learning rate decay of 0.1 after 210 epochs, while BeamsNetV2 was trained for 300 epochs with a learning rate decay of 0.1 after 255 epochs.
\begin{table}[h!]
\centering
\caption{Layer configuration and hyperparameters for both BeamsNet architectures: BeamsNetV1 and BeamsNetV2.}
\resizebox{\columnwidth}{!}{%
\begin{tabular}{|c|c|c|c|c|}
\hline
\textbf{Network} & \textbf{Layer} & \textbf{Input/Output} & \textbf{Kernel Size} & \textbf{Stride} \\ \hline
\multirow{5}{*}{BeamsNetV1} & Layer1 - IMU  & 6 → 9 & 100 & 20 \\ \cline{2-5} 
& Layer1 - DVL & 3 → 6 & 3 & 1 \\ \cline{2-5} 
& Layer2 - FC & 156 → 16 & - & - \\ \cline{2-5} 
& Layer3 - FC & 16 → 4 & - & - \\ \cline{2-5}
& Layer4 - FC & 12 → 3 & - & - \\ \hline
\multirow{4}{*}{BeamsNetV2} & Layer1 - DVL & 3 → 6 & 2 & 2 \\ \cline{2-5} 
& Layer2 - FC & 12 → 16 & - & - \\ \cline{2-5} 
& Layer3 - FC & 16 → 4 & - & - \\ \cline{2-5}
& Layer4 - FC & 12 → 3 & - & - \\ \hline
\end{tabular}%
}
\label{tab:1}
\end{table}

\subsection{Spectrally Normalized Memory Neuron Network (SNMNN)}
\begin{figure*}[h!]
	\centering
		\includegraphics[scale=0.25]{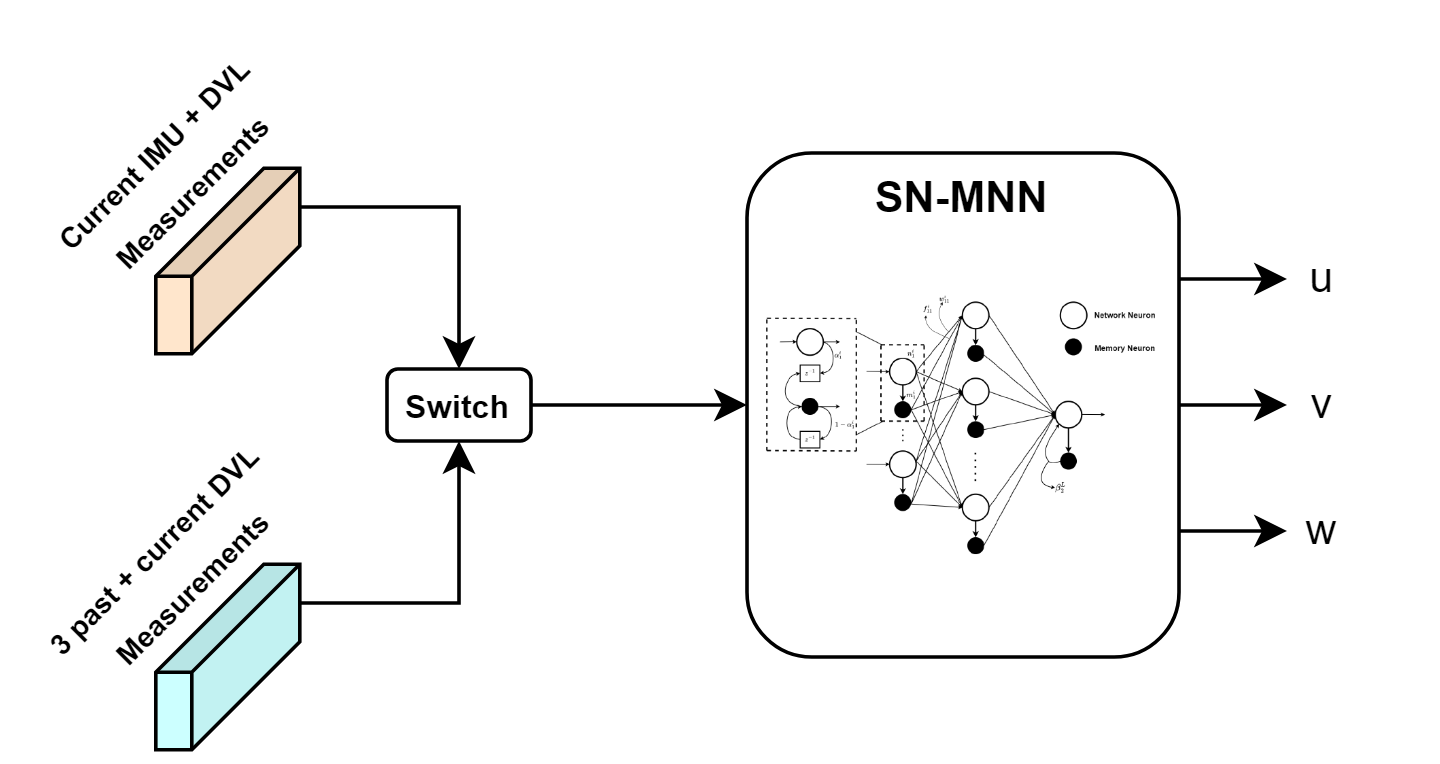}
	  \caption{The modified SNMNN architectures: SNMNNV1 and SNMNNV2}\label{fig2}
\end{figure*}

The SNMNN architecture is depicted in Fig. \ref{fig2}. It consists of three layers: input, one hidden and the output layer. The output layer predicts the AUV’s velocity vector in the body frame \cite{Makam2024}. There are two frameworks of the SNMNN examined. In the first version, current DVL beam measurements and IMU data are the inputs while the second version takes only current and three past instances’ DVL beam measurements as input assuming that no IMU data is available. The switch selects either the current IMU + DVL measurements or the 3 past + current DVL measurements as input into the SNMNN. 

The SNMNN uniquely combines fully connected network neurons with memory neurons, as described by \cite{sastry1994}. Each network neuron is paired with a memory neuron, incorporating memory elements to store and retrieve temporal information efficiently \cite{samanta2020}. During backpropagation, the weights of both regular and memory neurons are dynamically updated, enhancing the model's capacity to utilize historical data and improve predictive accuracy. The strength of SNMNNs lies in their ability to maintain stable predictions across various environments. To ensure reliable performance, spectral weight normalization is applied to both the network neuron weights ($W_{NN}$) and memory neuron weights ($W_{MN}$) \cite{Makam2024, rao2022}:
    \begin{equation}
        \overline{W}_{NN} = \left( \frac{W_{NN}}{\rho\left(W_{NN}\right)} \right) \gamma^{\frac{1}{L}} \label{W_bar}
       \end{equation}
    \begin{equation}
        \overline{W}_{MN} = \left( \frac{W_{MN}}{\rho\left(W_{MN}\right)} \right) \gamma^{\frac{1}{L}} \label{F_bar}
     \end{equation}
where $L$ is the layer number, $\gamma$ is the Lipschitz constant of the network, and $\rho(\cdot)$ denotes the spectral norm of the matrix.

The first version of the SNMNN architecture consists of 10 input neurons, one layer of 50 hidden neurons, and 3 output neurons. The network is trained for 50 epochs, maintaining a consistent learning rate set at 0.006 for regular learning and a memory coefficient learning rate set at 0.092. The second version consists of 16 input neurons, one layer of 60 hidden neurons, and 3 output neurons. The network is trained for 60 epochs, with the learning rates remaining unchanged from the first version. In both the architectures, the hidden layer is followed by a $tanh$ activation function and the output layer has a linear activation function.

\section{Results and Discussion}\label{RAD}

\subsection{Data Acquisition}

Field experiments were carried out in the Mediterranean Sea using the Snapir AUV to collect data. The Snapir AUV, a modified version of the A18D from ECA Group, is designed for deep-water operations and is capable of conducting autonomous missions at depths of up to 3000 meters. The AUV is equipped with an iXblue Phins Subsea INS, which utilizes fiber-optic gyroscope (FOG) technology for high-precision underwater navigation, and a Teledyne RDI Workhorse Navigator DVL. The INS operates at 100 Hz, while the DVL functions at 1 Hz.

The dataset consists of thirteen missions, each lasting 400 seconds, featuring various maneuvers, speeds, depths, and other conditions. Eleven of these missions were used for training the networks, while two were reserved for testing. The data is publicly available and thoroughly described in \cite{cohen2024kit}. The proposed approaches aim to enhance the accuracy of the AUV velocity estimation. To evaluate this, we implemented the error model outlined in \eqref{eqn:4}, incorporating an in-run bias of 0.001 m/s for each axis and a standard deviation of 0.15 m/s for the white noise. These corrupted beam velocity measurements were used as inputs to the networks, while the outputs represent the true AUV velocity, free from the added errors.
\begin{figure*}[h!]
\centering
\begin{tabular}{cc}
         \includegraphics[width=3.5in, height = 2.5in]{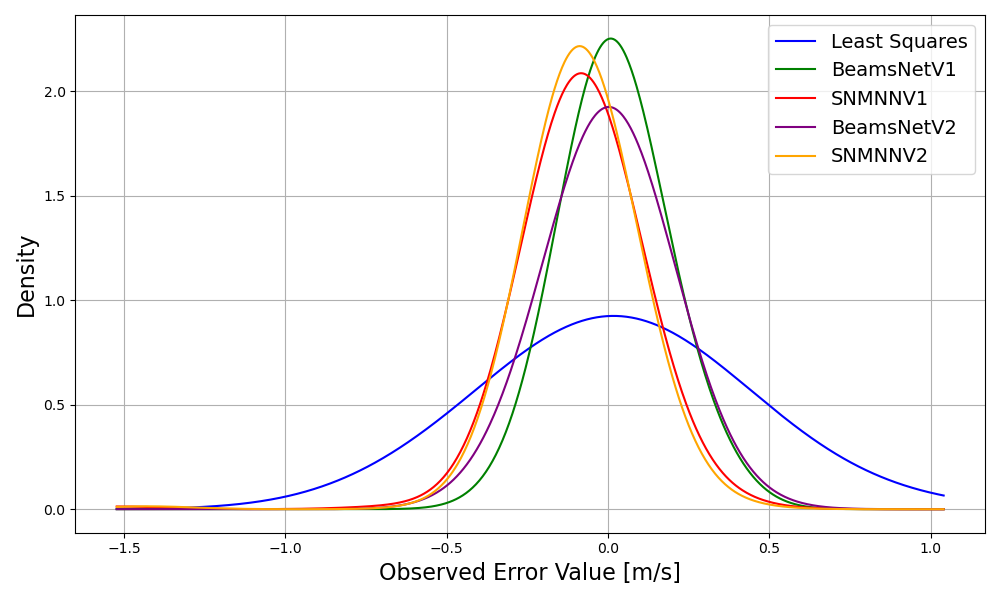}  &
        \includegraphics[width=3.5in, height = 2.5in]{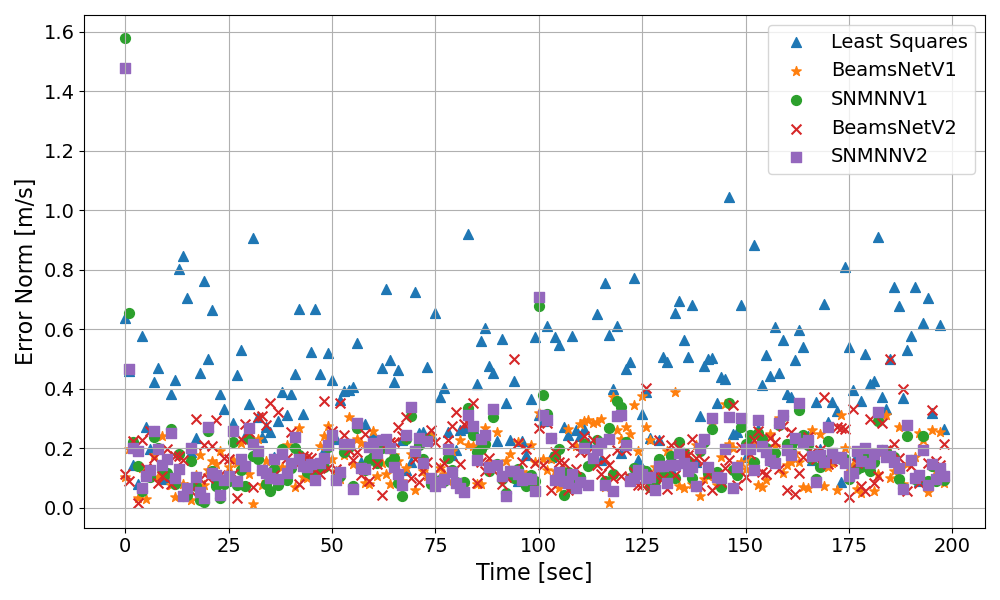} \\
        (a) & (b)
\end{tabular} 
  \caption{(a) The error norm density is given by the kernel density estimation (KDE) approach. (b) The error norm of the suggested approaches and the ground 
 truth. }
 \label{fig:3}
\end{figure*}

\subsection{Evaluation Metrics}
To assess the effectiveness of our proposed approach, we employ four performance metrics \cite{cohen2022}:
\begin{enumerate}
    \item Root Mean Squared Error (RMSE)\
    \begin{equation}
        RMSE(y, \hat{y}) = \sqrt{\frac{\sum_{i=0}^{N - 1} \left(y_i - \hat{y}_i\right)^2}{N}}. \label{rmse}    
    \end{equation}
    \item Mean Absolute Error (MAE)\
    \begin{equation}
        MAE(y, \hat{y})  = {\frac{\sum_{i=0}^{N - 1} \left|y_i - \hat{y}_i\right|}{N}}. \label{MSE}
    \end{equation}
    \item The coefficient of determination $R^2$\
    \begin{equation}
        R^2(y, \hat{y}) = 1 - \frac{\sum_{i=1}^{N} \left(y_i - \hat{y}_i\right)^2}{\sum_{i=1}^{N} \left(y_i - \bar{y}\right)^2}. \label{rsquare}
    \end{equation}
    \item Variance Accounted For (VAF) 
    \begin{equation}
        {VAF}(y, \hat{y}) = 100 \times \left[1 - \frac{var(y_i - \hat{y}_i)}{var(y_i)}\right]. \label{vaf}
    \end{equation}
\end{enumerate}

The RMSE and the MAE serve as metrics to quantify velocity errors, measured in units of [m/s], while $R^2$ and VAF are dimensionless. 

\subsection{Results}
After the networks were trained on the collected data, they were tested on unseen and different trajectories.
Table \ref{Table2} illustrates the test results for the LS solution and both versions of the SNMNN and BeamsNet architectures. The LS solution gives an overall RMSE of 0.2552 m/s, an MAE of 0.1855 m/s, an $R^2$ value of 0.9301 and a VAF of 93.02\%. 
Both BeamsNet (V1 and V2) and SNMNN (V1 and V2) demonstrated substantial improvements over the LS method in handling noisy and biased data. BeamsNetV1 reduced RMSE by 59.59\% and MAE by 58.76\%, while BeamsNetV2 achieved a 55.36\% RMSE reduction and a 55.07\% decrease in MAE. Similarly, SNMNNV1 improved RMSE by 56.07\% and MAE by 60.33\%, and SNMNNV2 reduced RMSE by 58.5\% and MAE by 59.77\%. Detailed performance metrics are provided in Table \ref{Table2}.

In terms of variance metrics, both models outperformed LS, which had a VAF of 93.02 and an $R^2$ of 0.9301. BeamsNetV1 achieved a VAF of 98.91 and an $R^2$ of 0.9886, reflecting significant improvements in predictive accuracy. BeamsNetV2 reached a VAF of 98.62 and an $R^2$ of 0.9861. Similarly, SNMNNV1 and SNMNNV2 attained VAF values of 98.66 and 98.80, with $R^2$ scores of 0.9865 and 0.9880, respectively. These higher VAF and $R^2$ values indicate a more precise fit and stronger predictive reliability under noisy and biased conditions compared to LS.



To demonstrate the enhanced velocity estimation achieved through learning-based approaches, Fig. \ref{fig:3}(a) presents a density error graph of the velocity norm. The graph was calculated using the kernel density estimation (KDE) approach \cite{chen2017tutorial} shows that while the mean of all methods slightly diverges from zero, the proposed approaches achieve a smaller variance, indicating greater certainty in the estimation. Additionally, Fig. \ref{fig:3}(b) presents the error in the velocities norm between the estimated velocity of each method and the ground truth, further illustrating the superiority of the learning-based approaches over the model-based one. Both the learning strategies clearly outperform the LS method, especially in handling noise and bias, making them more robust for predicting AUV velocities.

Although both BeamsNet and SNMNN architectures produced similar and superior results compared to the LS, it is important to emphasize that they operate based on different underlying concepts. The BeamsNet architecture relies on capturing spatial features from the input time-series data, and thus, its input must be constructed using a time window of measurements. As a result, the required storage is significantly higher compared to traditional approaches and the SNMNN. However, this allows the network to remain shallow with relatively few parameters. In contrast, SNMNN excels by capturing temporal dynamics through its memory units, enabling it to use only the current measurements as input without needing a time window. This makes SNMNN more efficient in terms of storage, while still achieving high predictive accuracy. Additionally, SNMNN is computationally lighter, making it a more scalable and practical choice for real-time applications, particularly in resource-constrained environments like AUV navigation.

\begin{table}[h!]
 \caption{Comparison of LS, BeamsNet and SNMNN}
\renewcommand{\arraystretch}{1.2}
\setlength{\tabcolsep}{5pt}
\centering
    \begin{tabular}{|c|c|c|c|c|} \hline 
         Metrics & RMSE [m/s] ($\downarrow$) & MAE [m/s] ($\downarrow$) &  VAF ($\uparrow$)  & $R^2$ ($\uparrow$)   \\ \hline \hline
         Least Squares & 0.2552  & 0.1855 & 93.02 & 0.9301 \\ \hline 
         BeamsNetV1    & 0.1031 & 0.0766  & 98.91 &0.9886\\ \hline 
         SNMNNV1       & 0.1121 & 0.0736 & 98.66 & 0.9865\\ \hline  
         BeamsNetV2    &0.1138  &0.0834  & 98.62 &0.9861\\ \hline   
         SNMNNV2       & 0.1059 & 0.0746 & 98.80 & 0.9880\\ \hline  
    \end{tabular}    \label{Table2}
\end{table}
\section{Conclusions}\label{con}
In this work, we present two learning-based alternatives to the model-based LS method for more accurate AUV velocity estimation. Tested on real AUV data, both spatial learning with BeamsNet and temporal learning with SNMNN showed significant improvements. Additionally, the velocity regression was more precise, with a smaller standard deviation than the LS method, suggesting that LS may not be the minimum variance estimator. These data-driven approaches are shallow, with fewer parameters, making them suitable for real-time applications.





\balance


\end{document}